\documentclass[11pt]{article}
\usepackage{epsfig}
\usepackage{a4wide}
\usepackage{amssymb}
\begin{document}

\title{Hubble operator in isotropic loop quantum cosmology}
\author{Golam Mortuza Hossain\\ 
{\it The Institute of Mathematical Sciences, Chennai 600 113,}\\
{\it India}\\
{\tt email:golam@imsc.res.in}}
\maketitle

\begin{abstract}
We present a construction of the Hubble operator for the spatially
flat isotropic
loop quantum cosmology. This operator is a Dirac observable on
a {\it subspace} of the space of physical solutions. This subspace gets selected
dynamically,  requiring that its action be invariant on the physical 
solution space. As a simple illustrative application of the expectation value
of the operator, we do find  a generic phase of 
(super)inflation, a feature shown by Bojowald \cite{minflation} 
from the analysis of effective Friedmann equation of loop quantum cosmology.
\end{abstract}

Pre-print No. IMSc/2003/06/11 \hfill

\section{Introduction}

	Symmetry reduction~\cite{msymmetry,mkinematics} at the quantum 
level of {\it loop quantum gravity}~\cite{rovellireview,thiemannreview,
thiemannlecture} has been applied to the field of cosmology with
notable success. The key result of this approach
{\it i.e. loop quantum cosmology} is the resolution of classical  
cosmological singularity~\cite{msingularity}. Apart from this, the 
consideration of effective Friedmann Equation due to small volume
modification of kinetic term of the matter fields leads to 
a phase of super-inflation with {\it graceful exit}~\cite{minflation}.
But there are lots of major issues yet to be clarified.
One such issue is the  
construction of physical observables~\cite{smolinoqg,rovellioqg} in
the loop quantum gravity.
This problem stems directly from the fact that defining physical
observables in general relativity itself is nontrivial as the physical 
observables are required to be gauge invariant objects. Consequently
in the quantum theory the corresponding operators are required to commute
with all the constraints of theory at least on the space of physical
solutions. But the problems of the quantum theory 
in this regards remains as hard as of the classical theory if not more.
Naturally this shortcoming is reflected in the loop quantum cosmology.
But in the case of loop quantum cosmology one works with a much simpler
setting compared to that of full theory. 
In particular only the Hamiltonian constraint needs to be addressed.
So one expects that this
problem will be easier to handle in the case of loop quantum cosmology.

	In this paper, from a general perspective, we emphasize
that a Dirac observable of a dynamical theory, to be physically relevant,
must be {\it localized} \cite{rovelli} in an appropriate frame of reference.
In the case of isotropic loop quantum cosmology, we show that 
the {\it localization} is no longer {\it an} option for construction of
Dirac observables rather it is the {\it only} option. In other words,
given a kinematical operator, one cannot construct the corresponding Dirac 
observable in the isotropic loop quantum cosmology without
{\it localization}. Further we show that the construction
of a Dirac observable allows one to choose only a {\it subspace} of the
space of physical solutions to be the domain of the corresponding 
operator. In this paper we will always refer constructed physical
observables as Dirac observables even though they do {\it not} 
act on the full physical solutions space.

	The motivation for the construction of the Hubble operator comes 
from the fact that Wheeler-DeWitt equation of minisuperspace quantum
cosmology for the FRW-deSitter universe can be written as an eigenvalue
equation of an operator which is classically proportional to the square
of the Hubble parameter. Importantly this operator involves the inverse
power of the {\it scale factor}. In isotropic loop quantum cosmology
there exists well defined, bounded operators for the inverse powers of
scale factor and they are constructed using the techniques which are 
used in the quantization of the full theory \cite{thiemannqsdV}. These 
operators have completely different spectrum 
compared to the classical values near the classical singularity.
So as one expects, we find that the expectation value of the Hubble operator 
constructed for the spatially flat isotropic loop quantum cosmology
behaves in a completely different way for a period of evolution near the 
classical singularity and this
period can be interpreted as a phase of super-inflation.

	In the section $2$ of this paper, we review the {\it must have}
properties of the physical observables in the context of both the
classical theory and the quantum theory. In section $3$, we consider the 
classical flat FRW universe. In the next section (section $4$) we 
recall \cite{kodama}
the Hamiltonian of the flat FRW universe in terms of
Ashtekar connection and densitized triad and then perform minisuperspace 
quantization. In the same  section we give a construction 
for the Hubble operator as a Dirac observable of this model.

In the beginning of the section $5$, we briefly review the kinematics of 
the isotropic loop quantum cosmology.  Then we study the properties of the
physical solution space and construct a class of {\it localized} Dirac 
observables for the isotropic flat loop quantum cosmology. We use this
construction to define the Hubble operator using the expression of the Hubble
operator defined in the context of minisuperspace quantization. For evaluating
the physical expectation value of the operator we define a physical
inner product which is motivated from the reformulation of ordinary 
quantum mechanics as constrained dynamics. Considering two simple examples we
evaluate the expectation value of the Hubble operator explicitly and then
discuss its generic properties near the classical singularity.

\section{Dirac Observables and Localization}

To understand the physical consequences of a dynamical theory, it is
absolutely necessary to figure out what are the physical observables
of the theory. Here we recall some of the key observations by Gaul and
Rovelli \cite{rovelli}. 

Let us consider a classical dynamical system which has gauge symmetry
in its basic variables. For such a system, given a set of initial
conditions, it is not possible for equation of motion to
uniquely specify the evolution of its basic variables.
Let us consider two solutions of equation of motion say $\phi(t)$
and $\phi'(t)$ which evolve from the same set of initial conditions
but get separated at some later time. In other words
$\phi(t_0) = \phi'(t_0)$ but  $\phi(t) \neq \phi'(t)$
for $ t > t_0$ where $t_0$ is the initial time. 
It was argued by Dirac that for such a case $\phi(t)$ and $\phi'(t)$
must be regarded as physically indistinguishable. If this is  not the case
then the classical {\it determinism principle} will fail to be true.

To avoid such a circumstances, the physical observables of a
theory are defined to be only those functions $\mathcal{O}(\phi(t))$
of basic variables which do not distinguish between $\phi(t)$ 
and $\phi'(t)$ {\it i.e.} $\mathcal{O}(\phi(t)) = \mathcal{O}(\phi'(t))$. 
As it was first systematically defined \cite{dirac} by Dirac they 
are also called Dirac observables. 
Thus it is clear that dynamical predictions only about
Dirac observables can be physically relevant.

Now one may ask whether all physical quantities that we can
measure  are necessarily Dirac observables? The answer is
{\it no}. To clarify this let us consider the example of a
simple pendulum. The classical theoretical setup contains
a deflection angle $\alpha$, a clock measuring time $t$
and a dynamical law namely the Newton's Law. Given a set of initial
conditions, from the dynamical law it is possible to uniquely predict
the deflection angle $\alpha$ at any time $t$ namely $\alpha(t)$.
So $\alpha(t)$ is a Dirac observable of the theory. However
the time $t$ itself cannot be {\it predicted} from the theory.
This implies that the time $t$, though we can measure it within 
the same setup, is not a Dirac observable of the theory. 
Another way to say it is that the time $t$ is a {\it parameter}
of the theory {\it not} a dynamical variable. But surely without
the association of time, the deflection angle $\alpha$ alone 
cannot describe the dynamics of the system.

Following Gaul and
Rovelli \cite{rovelli} let us introduce the notion of {\it partial
observables} (see \cite{rovellipartial} for more details).
For the pendulum system, time $t$ is an {\it independent
partial observable} and deflection angle $\alpha$ is a {\it dependent
partial observable}. Partial observables individually are not physically
meaningful quantities of the theory unless they are combined to form
the {\it complete observables} like $\alpha(t)$.
So to be physically relevant, a Dirac observable must be represented
by a {\it complete observable}.
Formally Dirac observables of a classical theory with constraints 
are defined to be only those functions of the phase space variables whose
Poisson bracket vanishes atleast weakly with all the first class constraints.

The essential change that is required when we move into Einsteinian
framework from Newtonian framework, is the modification 
in the notion of reference system.
In the Newtonian framework there is a fixed space and a fixed time. 
These objects define the {\it frame of reference} for a dynamical
system. Most importantly the frame of reference is completely
independent of the dynamics of the system.
Now the frame of reference defines the independent partial
observables whereas dynamical variables of the system define the
dependent partial observables.
In construction of a complete observable for the system, we need to 
associate
independent partial observables with the dependent partial observable.
Following Gaul and Rovelli \cite{rovelli}, let us call this process
the {\it localization} of the dynamical variable in the frame of reference.
Only {\it localized} objects can be the candidates for the physically 
relevant Dirac observables of the theory.

	In general relativity situations are very different from that
of Newtonian framework. In a generally covariant theory,  
the puzzle of {\it hole argument}
(see \cite{rovelli} and reference therein) and its resolution imply that
it is no longer possible to separate the frame of reference from the 
dynamics of the system. So in general relativity we can no longer
distinguish an independent
partial observable from a dependent partial observable.
As a consequence the physical observables in general relativity are
defined only with respect to each other. In other words for the
purpose of localization which is necessary to define a complete
observable, we must use some of the degrees of freedom of the theory
either from the matter sector or from the gravitational sector.

	In the canonical quantization of a classical theory the
classical observables are represented by operators on an appropriate
Hilbert space. For Dirac observables, the requirement of vanishing
Poisson bracket is turned into the requirement of vanishing 
commutator bracket with all the constraints atleast on the space of 
physical solutions.

From a quantum
theory one obtains the physically relevant quantities from the
expectation values of the operators in the physical states. So it is naturally
expected that in a quantum theory of general relativity the 
Dirac observables and its expectation value should incorporate
the notion of classical localization in an appropriate way.

	In the case of isotropic loop quantum cosmology we
will see that for a class of operators the requirement of 
vanishing commutator with the Hamiltonian constraint on the physical solution
space itself provides a natural prescription for the localization of 
the operators.

\section{Classical FRW Cosmology}

In general relativity, the dynamics of gravity manifest itself through the
dynamics of spacetime. 
A homogeneous and isotropic spacetime is described by the 
Friedmann-Robertson-Walker (FRW) metric. For the spatially flat FRW
spacetime the invariant line element, in natural units($c=\hbar=1$),
is given by
\begin{equation}
ds^2  ~=~  -dt^2 + a^2(t) {\bf dx}^2 ~,
\label{frw_metric}
\end{equation}
where $a(t)$ is the {\it scale factor} of the FRW universe.

The dynamics of the spacetime (\ref{frw_metric}) is described by
the famous Friedmann equation 
\begin{equation}
3~{\left({\dot a \over a}\right)}^2 ~=~ k~\rho ~+~ \Lambda ~,
\label{fe1}
\end{equation}
where overdot denotes the derivative with respect to coordinate 
time $t$, $k := 8\pi G$ is the gravitational coupling constant, 
$\rho$ is the homogeneous energy density of the matter fields and
$\Lambda$ is the cosmological constant.
Friedmann equation (\ref{fe1}) along with covariant conservation
of matter stress tensor completely determine the evolution 
of FRW universe upto the boundary conditions.

We now consider the simplest example of a FRW universe where
the dynamics is completely controlled by a non zero cosmological
constant. This universe is also known as deSitter universe.
The Friedmann equation (\ref{fe1}) for the FRW-deSitter case leads
to
\begin{equation}
\mathcal{H}^2 ~=~ {\Lambda \over 3},
\label{fe3}
\end{equation}
where the quantity $\mathcal{H}:=\left({\dot a \over a}\right)$ is the 
Hubble parameter. One should notice here that the Hubble parameter 
for FRW-deSitter universe is a constant.

Throughout this paper we will consider FRW-deSitter universe as a test
case for comparing results of classical cosmology, minisuperspace
(Wheeler-DeWitt) quantum cosmology and loop quantum cosmology.

\section{Minisuperspace (Wheeler-DeWitt) Quantum Cosmology}

In minisuperspace quantization~\cite{dewitt,misner} of
gravity, at first one reduces the
classical phase space of general relativity by the symmetry of the
spacetime. Thereafter one uses the symmetry reduced phase space
variables for the quantization procedure.

The Einstein-Hilbert action for the spatially flat FRW spacetime 
(\ref{frw_metric}) can be written as 
\begin{equation}
S_{grav}  ~:=~ \int dt L_{grav},
\label{action_lg}
\end{equation}
where
\begin{equation}
L_{grav}  ~=~  - {~3~ a ~{\dot a}^2 \over k}~-~{\Lambda \over k} ~a^3 ~.
\label{lg}
\end{equation}
In getting expression(\ref{lg}) for $L_{grav}$ we have dropped a total
derivative term. Moreover to avoid infinity due to integration over whole
space, as integrand is independent of position, we have compactified 
the space and taken the coordinate volume integral to be one.

	The spatial part of the FRW metric is given
by $g_{ab} ~=~ a^2 ~\delta_{ab}$. So the metric has just one degree
of freedom. For simplicity we now define a new {\it position}
variable $Q ~:=~ a^2$. The corresponding conjugate {\it momentum} is given by
$P ~:=~ {\partial L_{grav} \over {\partial {\dot Q}}} $. Then the 
Hamiltonian for the spacetime is given by
\begin{equation}
H_{grav} ~:=~ P~{\dot Q} ~-~ L_{grav} 
 ~=~ -{k \over 3}~ P^2 {\sqrt Q} ~+~{\Lambda \over k}~ Q^{3\over 2} ~.
\label{ham_g1}
\end{equation}

Now $Q$ and $P$ being a canonical pair, we can perform minisuperspace
quantization \cite{dewitt,misner} using them. However as we intend to
import few ideas into loop quantum cosmology, we define a new set
of canonical pair as
\begin{eqnarray}
c ~&:=&~ \gamma {\dot a}  ~=~ - ~{1 \over 3}~\gamma k~P ~,  
\label{can_pairc} \\
p ~&:=&~ a^2 ~=~ Q ~,
\label{can_pairp}
\end{eqnarray}
where $\gamma$ is a positive real number. In the context of loop quantum
gravity $\gamma$ is known as {\it Barbero-Immirzi parameter}
\cite{barbero,immirzi} and 
the new canonical pair $c$ and $p$ are the Ashtekar connection and 
the densitized triad for the homogeneous and isotropic
spacetime. It is easy to check that the Poisson bracket between these new
variables is given by
\begin{equation}
\{c,p\} ~=~  ~{1 \over 3}~\gamma k~.
\label{poisson}
\end{equation}

Including the matter sector and using the new pair of variables $(c,p)$ 
the full Hamiltonian can be written as 
\begin{equation}
H ~:=~ H_{grav} ~+~ H_{matter} 
	 ~=~ -{3 \over {k\gamma^2}}~ c^2 {\sqrt p}
	 ~+~{\Lambda \over k}~ p^{3\over 2} ~+~ H_{matter}~.
\label{ham_full1}
\end{equation}

In performing canonical quantization, one represent the 
classical phase space variables as operators on an appropriate
Hilbert space and their classical Poisson bracket is turned into
a commuter bracket for the operators. Here it is convenient to choose
triad representation for the canonical quantization. In this
representation triad acts as a multiplication operator whereas
the connection acts as a derivative operator which is given by
\begin{equation}
\hat{c} ~=~  {i\over 3} \gamma k {d\over dp}  ~.
\label{con_op}
\end{equation}

Since $\hat{c}$, $\hat{p}$ do not commute with each other, the ordering
ambiguity arises naturally in constructing the Hamiltonian operator.
Here we will choose the {\it connection to the left} ordering as this 
appears as an approximation of Wheeler-DeWitt operator \cite{mwdw}
of isotropic loop quantum cosmology~\cite{misotropic}.

In the quantum theory the Hamiltonian acts as a constraint operator
on the kinematical Hilbert space, because in the classical general
relativity the corresponding Hamiltonian vanishes. The physical
states $\psi(p)$ of the quantum theory are selected 
out by the requirement that the Hamiltonian constraint (\ref{ham_full1})
annihilate the physical states {\it i.e.}
\begin{equation}
\hat{H} ~\psi(p) ~=~ 0 ~.
\label{ham_cons1}
\end{equation}

To solve the equation (\ref{ham_cons1})
explicitly we need to specify the matter sector Hamiltonian $H_{matter}$.
Here again we consider the FRW-deSitter universe. For this case the
matter sector Hamiltonian is zero. Using the expression for derivative
operator $\hat{c}$, the Hamiltonian 
constraint leads to a second order linear differential equation which
is given by
\begin{equation}
{d^2\over {dp^2}} \left( {\sqrt p} ~\psi(p)\right) 
~+~ b^2~ p^{3\over 2} \psi(p) ~=~ 0 ~,
\label{ham_cons2}
\end{equation}
where $b^2 = {3~\Lambda \over k^2}$. The solutions of the equation 
(\ref{ham_cons2}) are given by Bessel functions of fractional order
$1\over3$ and $-{1\over3}$ {\it i.e.}
\begin{equation}
\psi_{\pm}(p)~=~ J_{\pm \frac{1}{3}}({2\over 3} b~ p^{3\over 2}) ~.
\label{sol1}
\end{equation}

Physical observables of a theory are represented by 
Dirac observables.
In the quantum theory it amounts to the requirement
that the corresponding operator $\hat{O}$ must commute with the Hamiltonian
constraint at least on the space of physical states {\it i.e.}
\begin{equation}
[\hat{O},\hat{H}] ~\psi(p) ~=~ 0 ~.
\label{ms_dirac_op}
\end{equation}

From the equation (\ref{ham_cons2}), we observe that the 
$\psi_{\pm}(p)$ appear as the ``eigen functions'' of the operator
$ p^{-{3\over 2}}{d^2\over {dp^2}} {\sqrt p}$. This operator classically
corresponds to $- {9\over{\gamma^2 ~k^2}}{c^2 \over p} 
~=~ -{9 \over k^2} {\left({\dot a \over a} \right)}^2 $. 
Most importantly the classical quantity is
proportional to the square of the Hubble parameter. Observing
this fact, we can now define the formal Hubble operator for this model as
\begin{equation}
\hat{\mathcal{H}^2} ~=~ -{k^2 \over 9}~ p^{-{3\over 2}}
	{d^2\over {dp^2}} {\sqrt p} ~.
\label{ms_hubble_op}
\end{equation}

$\psi_{\pm}(p)$ being eigen functions, $\hat{\mathcal{H}^2}$ trivially
satisfy (\ref{ms_dirac_op}) and so $\hat{\mathcal{H}^2}$ is a Dirac observable.
It is possible to choose a different operator ordering other than 
(\ref{ms_hubble_op}) such that the new operator also corresponds
classically to the square of Hubble parameter.
However all such operators fail to commute 
with the Hamiltonian constraint on the physical solutions space.
The formal expectation value of this operator in the physical states,
with respect to the kinematical inner product, is given by
\begin{equation}
\langle \hat{\mathcal{H}^2} \rangle ~:=~ 
{{ \langle \psi_{\pm}(p), \hat{\mathcal{H}^2} \psi_{\pm}(p) \rangle} \over 
{ \langle \psi_{\pm}(p), \psi_{\pm}(p) \rangle}} \\
~=~ {\Lambda \over 3}.
\label{ms_hubble_exp}
\end{equation}

So we find that the expectation value of the Hubble operator 
$\hat{\mathcal{H}^2}$ is exactly equal to the classical value (\ref{fe3}).
In other words, in the example considered here,
the minisuperspace quantization does not give 
any correction to classical Friedmann equation (\ref{fe3}).
However it is important to emphasize here that, for a general constrained 
system, the definition of expectation value (\ref{ms_hubble_exp}) is 
not always a well defined quantity as the solutions for such system
are not always square-integrable functions.

\section{Isotropic Loop Quantum Cosmology}
	
	The isotropic loop quantum cosmology~\cite{misotropic} 
is defined to be the sector of full loop quantum gravity whose
kinematical~\cite{mkinematics} distributional states are supported only on  
homogeneous and isotropic connections. Unlike
minisuperspace quantization, in loop quantum cosmology symmetry
reduction are done at the quantum level. 
So many features of the full theory like discrete volume spectrum
\cite{mvolume} survive the symmetry reduction.

\subsection{Kinematics and Dynamics}

	In this paper we will consider only spatially flat model.
We will follow mainly the conventions of \cite{mboundary,mmatter}. 
The symmetry reduced kinematical Hilbert space for the isotropic loop quantum
cosmology is given by $L^2(SU(2),d\mu_H)$. An orthonormal basis set for the
kinematical Hilbert space is given by 
$\{ |n\rangle ~|~ n \in {\mathbb Z}\}$. These
basis states are the eigenstates of the isotropic triad operator. Here we
{\it define} a new triad operator $\hat{p}$ whose eigenvalues differ
from that of \cite{mmatter} by a factor of $l_p^2$. The action of this
operator $\hat{p}$ on the triad basis states is defined as
\begin{equation}
\hat{p}| n \rangle ~=~ {1 \over 6}\gamma ~n ~|n\rangle .
\label{lqc_triad_op}
\end{equation}
One should notice here the eigenvalues of $\hat{p}$ are dimensionless.
So this operator can be related with the {\it scale factor} of the FRW
metric (\ref{frw_metric}) directly by the relation(\ref{can_pairp}). 

In the connection representation, the functional form of these basis states is
given by
\begin{equation}
\langle c|n \rangle ~=~ {{exp({1\over2} i n c)} 
			\over {\sqrt{2} ~sin({1\over 2} c)}} .
\label{lqc_basis_con}
\end{equation}

	The isotropic volume operator and the inverse scale factor operator
both are diagonal in these basis. The action of the volume operator on these
basis is given by
\begin{equation}
\hat{V}| n \rangle ~=~ ({1 \over 6}\gamma l_p^2)^{3\over2}
  \sqrt{(|n|-1)|n|(|n|+1)} ~|n\rangle 
		   ~=:~ V_{{1\over2}(|n|-1)} |n\rangle ~,
\label{lqc_volume_op}
\end{equation}
and the action of the inverse scale factor is given by
\begin{equation}
\hat{a^{-1}}| n \rangle ~=~ 16 \gamma^{-2} l_p^{-3}
\left(\sqrt{ V_{{1\over2}|n|}} - \sqrt{V_{{1\over2}|n|-1}}\right)^2 |n\rangle ~.
\label{lqc_isf_op}
\end{equation}
Here again we have defined the operator such that the eigenvalues of the 
inverse scale factor are dimensionless.

	A general kinematical state in these basis can be written as
\begin{equation}
|s\rangle ~=~ \sum_{n \in {\mathbb Z} } s_n ~|n\rangle ~,
\label{lqc_gen_state}
\end{equation}
where coefficients $s_n$'s are in fact vectors of the Hilbert space
corresponding to the matter sector. In coordinate representation of
conventional quantization of matter sector, $s_n$'s are complex
valued functions of the matter degrees of freedom.

	The full Hamiltonian constraint of loop quantum cosmology consists
of gravitational sector as well as matter sector. In standard dynamical
system, gravity couples with matter sector via metric components. In loop
quantum gravity metric variables are replaced by triads. So complete
geometrical properties are encoded in the triad components whereas its dynamics
is encoded in the connection components. Thus for standard gravity-matter
coupling, the matter sector Hamiltonian is diagonal in the triad basis.
	The action of gravitational Hamiltonian on the
triad basis is given by
\begin{equation}
\hat{H}_{grav} |n\rangle ~=~ {3\over2}\gamma^{-2} (k\gamma l_p^2)^{-1} 
	sgn(n)(V_{{1\over2}|n|} - V_{{1\over2}|n|-1})
	(|n+4\rangle -2|n\rangle +|n-4\rangle) ~, 
\label{lqc_ham_grav}
\end{equation}
and the action of matter Hamiltonian is given by
\begin{equation}
\hat{H}_{matter} |n\rangle ~=~ H_{matter}(n)~|n\rangle ~. 
\label{lqc_ham_matter}
\end{equation}
In (\ref{lqc_ham_matter}), $H_{matter}(n)$ is just the symbolic
eigenvalue of the matter Hamiltonian but one should remember that it is
still an operator on the Hilbert space of matter sector.

	To select out the physical states $|s\rangle$ one imposes 
Hamiltonian constraint as
\begin{equation}
\hat{H} ~|s\rangle ~:=~ (\hat{H}_{grav} ~+~ \hat{H}_{matter})|s\rangle ~=~ 0 ~.
\label{lqc_ham_cons}
\end{equation}
The combined action of $\hat{H}_{grav}$ and $\hat{H}_{matter}$ 
on the state $|s\rangle$ leads to
\begin{equation}
\sum_{n \in {\mathbb Z}}( A_{n+4}s_{n+4}~+~B_{n}s_{n} ~+~ A_{n-4}s_{n-4})
	|n\rangle ~=~ 0 ~,
\label{lqc_ham_cons2}
\end{equation}
where $A_n = V_{{1\over2}|n|} - V_{{1\over2}|n|-1}$ and 
$B_n = -2 A_n + {2\over3} \gamma^2 (k~\gamma l_p^2) H_{matter}(n)$. In
(\ref{lqc_ham_cons2}) we have absorbed $sgn(n)$ in the definition of $s_n$.
Now the equation (\ref{lqc_ham_cons2}) requires coefficient of all 
$|n\rangle$ to vanish. The coefficient of $|n+4\rangle$ leads to the
difference equation
\begin{equation}
A_{n+8}s_{n+8}~+~B_{n+4}s_{n+4} ~+~ A_{n}s_{n} ~=~ 0 ~.
\label{lqc_diff_eqn}
\end{equation}

	The difference equation(\ref{lqc_diff_eqn}) for $s_n$ can also
be regarded as an evolution equation \cite{mdiscrete} with respect
to an {\it internal time}.
There one chooses triad basis index $n$ as an internal time. $n$ being 
discrete, the evolution is necessarily discrete evolution.

\subsection{Physical Solution Space}

The equation (\ref{lqc_diff_eqn}) is an eighth order difference equation
but with equal steps of four. This implies that we can divide all solution 
$s_n$'s of (\ref{lqc_diff_eqn}) into four sector and each sector can be 
completely determined by specifying two initial conditions.

	We now define the set of all $s_n$ which are solution of 
(\ref{lqc_diff_eqn}) with the given initial conditions $(s_i, s_{4+i})$ as
\begin{equation}
G_i(s_i,s_{4+i}) ~:=~ \{s_{4n+i}~|~ n \in {\mathbb Z}\} ~,
\label{lqc_sol_set}
\end{equation}
where $i=0,1,2,3$.
The state vector corresponding to the set $G_i(s_i,s_{4+i})$ is given
by
\begin{equation}
|G_i\rangle_{(s_i,s_{4+i})} ~:=~ \sum_{n} s_{4n+i} |4n+i\rangle ~.
\label{lqc_sol_state1}
\end{equation}
So clearly any physical state $|s\rangle$ can be written as
\begin{equation}
|s\rangle ~=~ |G_0\rangle_{(s_0,s_4)} ~+~ |G_1\rangle_{(s_1,s_5)} 
	~+~|G_2\rangle_{(s_2,s_6)} ~+~|G_3\rangle_{(s_3,s_7)} ~.
\label{lqc_sol_state2}
\end{equation}

The state $|G_i\rangle_{(s_i,s_{4+i})}$'s are orthogonal to each other. Moreover
by construction the Hamiltonian constraint will annihilate them individually. 
Thus $|G_i\rangle_{(s_i,s_{4+i})}$'s are also physical states.

	To specify a general solution of (\ref{lqc_diff_eqn}) one requires
to supply eight initial conditions. But we have seen that we can decompose a
general solution into four orthogonal solutions each of which requires 
just two initial conditions and they cannot be decomposed further. 
We will see in the next section that this splitting is crucial for the
construction of a class of Dirac observables of isotropic loop quantum 
cosmology. Moreover for the following 
discussion it is sufficient to consider just one such sector say
$G_1 := \{~|G_1\rangle_{(s_1,s_5)}\}$. Clearly any physical state in this sector
can be uniquely constructed by specifying values of $s_1$ and $s_5$.

	If we now consider the $s_1$-$s_5$ plane (see fig. 1), corresponding
to every point in the plane there exist a physical state in the $G_1$ sector.

\subsection{Localized Dirac Observable}

Any function of classical phase space
variables of isotropic loop quantum cosmology can be either purely a function 
of triad or purely a function of connection or it could be a function of
both. For simplicity, we will consider here only the functions of the 
first category. The corresponding operators are diagonal in the triad basis.
We have already argued that for a dynamical system, the physically relevant
Dirac observables must be localized in an appropriate frame of reference.
In the following construction for the localized Dirac observable, 
we start with a kinematical operator and then explore the requirement on 
physical solution space such that the action of the kinematical operator 
commutes with the Hamiltonian constraint {\it at some fixed time slice}. Using
this requirement we show that it is not possible for a single Dirac observable
to represent the action of the kinematical operator on all the triad basis
vectors. Then we give a procedure for construction of Dirac observables for a 
given kinematical operator. Although we start with a kinematical operator
whose classical expression can be written as a pure function of triad variable,
at the end of the construction, the corresponding Dirac observables 
do not have this property {\it i.e.} they cannot be classically represented
by simple analytic functions of triad variable.

The action of an diagonal operator say $\hat{O}$, on the triad basis can 
be written as
\begin{equation}
\hat{O}~|n\rangle ~=~ f_n~|n\rangle ~.
\label{lqc_oc1}
\end{equation}
We now consider a general state $|s'\rangle$ which is spanned
by the basis set $\{|4n+1\rangle~|~n \in {\mathbb Z}\}$ as
\begin{equation}
|s'\rangle ~=~ \sum_{m} s_m' ~|m\rangle ~,
\label{lqc_ks1}
\end{equation}
where $m=4n+1$. 
The action of the operator $\hat{O}$ on the state $|s'\rangle$ is given by
\begin{equation}
\hat{O}~|s'\rangle ~=~ 
\sum_m s'_m f_m |m\rangle ~.
\label{lqc_ostate}
\end{equation}

The projection of the action of the Hamiltonian on the 
state $|s'\rangle$ at a fixed ``time slice'' say $(m_0+4)$, is
given by
\begin{equation}
\langle m_0+4|\hat{H}|s'\rangle ~=~ 
A_{m_0+8}s'_{m_0+8}~+~B_{m_0+4}s'_{m_0+4} ~+~ A_{m_0}s'_{m_0}~.
\label{lqc_proj1}
\end{equation}

If the state $|s'\rangle$ also belongs to the physical solution space 
then the coefficients $s'_{m_0+8}$, $s'_{m_0+4}$ and $s'_{m_0}$ will 
satisfy 
\begin{equation}
A_{m_0+8}s'_{m_0+8}~+~B_{m_0+4}s'_{m_0+4} ~+~ A_{m_0}s'_{m_0} ~=~ 0~.
\label{lqc_relation1}
\end{equation}
$m_0$ being a fixed time slice, the equation (\ref{lqc_relation1}) is just a
relation and {\it not} the difference equation.

Similarly the projection of the action of the Hamiltonian on the 
state $\hat{O}|s'\rangle$ at the same time slice $(m_0+4)$, is
given by
\begin{equation}
\langle m_0+4|\hat{H}~\hat{O}|s'\rangle ~=~ 
A_{m_0+8}s''_{m_0+8}~+~B_{m_0+4}s''_{m_0+4} ~+~ A_{m_0}s''_{m_0} ~,
\label{lqc_proj2}
\end{equation}
where $s''_{m_0+8}=f_{m_0+8}s'_{m_0+8}$, $s''_{m_0+4}=f_{m_0+4}s'_{m_0+4}$  
and $s''_{m_0}=f_{m_0}s'_{m_0}$.  
We now explore the requirement on $(s'_{m_0+8}, s'_{m_0+4}, s'_{m_0})$
such that $(s''_{m_0+8}, s''_{m_0+4}, s''_{m_0})$ 
also satisfy the same relation (\ref{lqc_relation1})
{\it i.e.}
\begin{equation}
A_{m_0+8}s''_{m_0+8}~+~B_{m_0+4}s''_{m_0+4} ~+~ A_{m_0}s''_{m_0} ~=~ 0~.
\label{lqc_relation3}
\end{equation}

The equations (\ref{lqc_relation1}) and 
(\ref{lqc_relation3}) together lead to a consistency requirement on 
 $s'_{m_0+4}$ and $s'_{m_0}$ and that is given by
\begin{equation}
(f_{m_0+8}-f_{m_0+4})~B_{m_0+4}s'_{m_0+4} 
	~+~ (f_{m_0+8} - f_{m_0})A_{m_0}s'_{m_0} ~=~ 0~.
\label{lqc_relation4}
\end{equation}
For the trivial case where the functions $f_{m_0}$'s are independent of $m_0$ 
the equation (\ref{lqc_relation4}) is empty. 
We can now interpret the relation (\ref{lqc_relation3})
also as a projection of the action of Hamiltonian on some physical state
$|s''\rangle$ at the same time slice $(m_0+4)$. 

	$|s'\rangle$ being a physical state, by specifying 
the values of $s'_{m_0+4}$ and $s'_{m_0}$, we can construct all other 
$s'_m$ using the difference equation (\ref{lqc_diff_eqn}).  Since the equation 
(\ref{lqc_diff_eqn}) is a linear difference equation then the consistency 
requirement (\ref{lqc_relation4}) will lead to a linear equation between 
$s'_1$ and $s'_5$. We write this linear equation as
\begin{equation}
s'_5  ~=~ \beta'( m_0+4)~ s'_1 ~.
\label{lqc_cond1}
\end{equation}

	Similarly starting from $s''_{m_0+4}$ and $s''_{m_0}$ one can
construct the whole tower of $s''_m$ using the difference
equation (\ref{lqc_diff_eqn}).
So by construction $|s''\rangle$ will be a physical state. We denote
the relation between $s''_1$ and $s''_5$ for $|s''\rangle$ as
\begin{equation}
s''_5  ~=~ \beta''( m_0+4)~ s''_1 ~.
\label{lqc_cond2}
\end{equation}

	In $s_1$-$s_5$ plane (see fig. 1) two physical states
$|s'\rangle$ and $|s''\rangle$ are described by a point on the straight 
lines of different slope. 
We now define a set of physical states of the $G_1$ sector as
$D_{m_0+4} := \{~|G_1\rangle_{(s_1,s_5)} ~|~ s_5 = \beta'(m_0+4)~s_1\}$.

\begin{center}
\epsfxsize=100mm
\epsfysize=70mm
\epsfbox{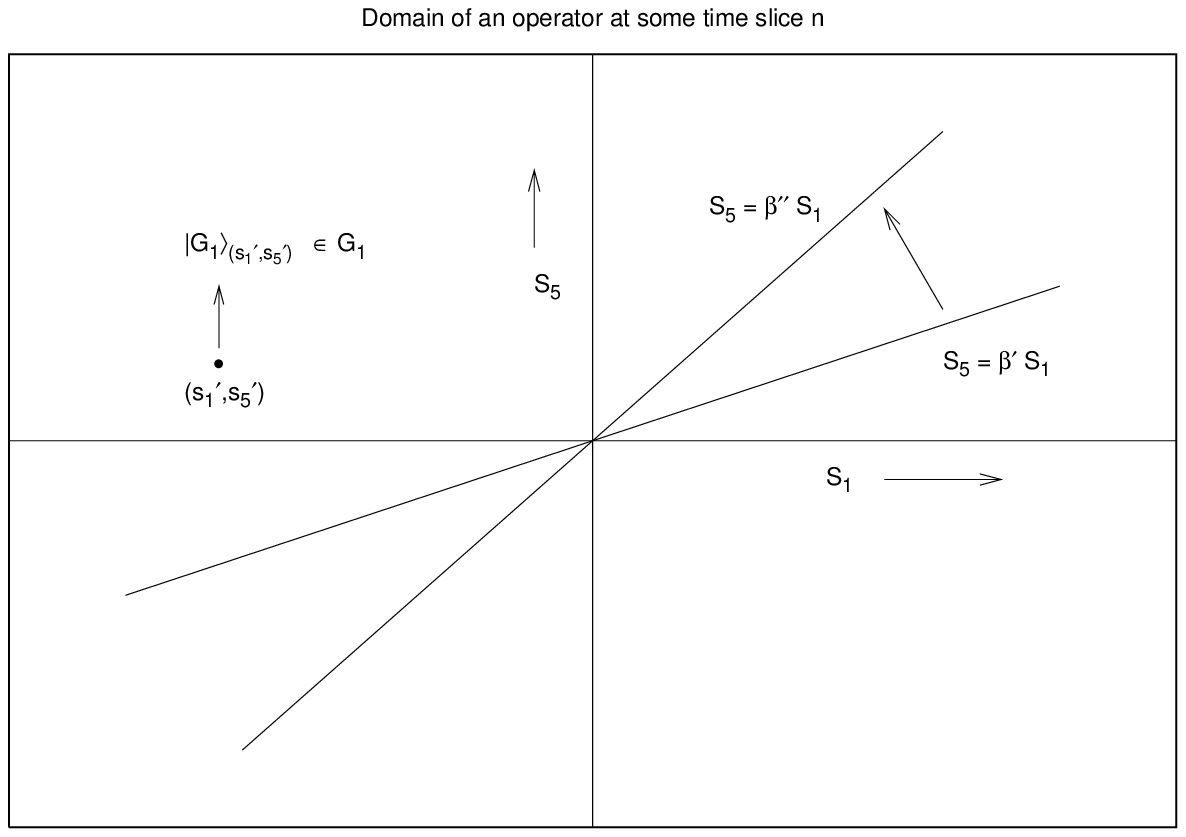}

{Figure 1. Domain of an operator at some time slice n.}
\end{center}

	Similar argument readily goes through for the $G_2$ and $G_3$
sector. But for $G_0$ sector equation (\ref{lqc_relation4}) leads to 
$s_4=0$. Consequently all $s_{4n}$ drop out except for $s_0$ which anyway
cannot be determined by (\ref{lqc_diff_eqn}) as it decouples from the rest
of $s_n$'s. We now generalize the definition of the subspaces as
$D_{4n+i} := \{~|G_i\rangle_{(s_i,s_{4+i})}~|~ s_{4+i} = \beta'(4 n + i)~s_i\}$.
For the $G_0$ sector the subspaces are trivial and they are given by
$D_{4n} = \{~|G_0\rangle_{(s_0,s_{4})} = s_0|0\rangle\}$. One may note
here that the subspaces exclude the pre-classical solutions.

For the operator $\hat{O}$ (\ref{lqc_oc1}) to be a Dirac observable one 
requires it to satisfy
\begin{equation}
\hat{H}~\hat{O}~|s\rangle ~=~ 0 ~,
\label{lqc_dirac}
\end{equation}
where $|s\rangle$ is a physical state. 
We use the decomposition (\ref{lqc_sol_state2}) of the physical state
$|s\rangle$. Then the equation (\ref{lqc_dirac}) implies  
$\hat{H}\hat{O}|G_i\rangle_{(s_i,s_{4+i})} ~=~ 0$, for all $i$. Here
we consider just the $G_1$ sector of the physical solution space.  
Inserting the complete set of basis from the left, we get the requirement
in the $G_1$ sector as
\begin{equation}
\sum_{n}|n\rangle \langle n| \hat{H}~\hat{O}~|G_1\rangle_{(s_1,s_5)}
~=~ \sum_{m}|m\rangle \langle m| \hat{H}~\hat{O}~|G_1\rangle_{(s_1,s_5)}
~=~ 0 ~,
\label{lqc_dirac2}
\end{equation}
where $m=4n+1$.

	The equation (\ref{lqc_dirac2}) requires the coefficient 
$\langle m| \hat{H} \hat{O}~|G_1\rangle_{(s_1,s_5)}$ to vanish for 
all time slice $m$. We have seen that this requirement, at a given time 
slice $m$, can be satisfied provided the operator $\hat{O}$ acts only on
a subspace $D_m$ of the $G_1$ sector of the physical solution space.
To ensure this we need to define the domain of $\hat{O}$ in $G_1$ sector
to be the subspace $D_m$.
Now the domain $D_m$ being $m$ dependent, we must tag the operator 
$\hat{O}$ with $m$ as well. This implies that to satisfy (\ref{lqc_dirac2})
for all $m$ we need to have a collection of operators each tagged with $m$.

	We now define the {\it tagged} operator $\hat{O}_m$ as
\begin{equation}
|s''\rangle ~=~ \hat{O}_m~|s'\rangle ~,
\label{lqc_to1}
\end{equation}
where $|s'\rangle \in D_m$ and $|s''\rangle$ is constructed using the difference
equation (\ref{lqc_diff_eqn}) with the given values of $s''_m = f_m s'_m$,
$s''_{m-4} = f_{m-4} s'_{m-4}$ and the operator $\hat{O}_m$ is assumed 
to be diagonal like the kinematical operator $\hat{O}$.
Using the definition (\ref{lqc_to1}), one can
uniquely construct the matrix elements of $\hat{O}_m$ in the triad basis of
$G_1$ sector and this is possible only because of the fact that 
the physical states split into four orthogonal states; each requiring
just two initial conditions.
This operator $\hat{O}_m$ has exactly the same action, as of 
the kinematical operator $\hat{O}$ (\ref{lqc_oc1}), on the three successive 
basis $|m-4\rangle$, $|m\rangle$ and $|m+4\rangle$ of the $G_1$ sector. 
Now we need to specify the matrix elements of $\hat{O}_m$ in $G_2$, $G_3$
and $G_0$ sectors. 
Since we have defined the operator $\hat{O}_m$ with respect to its nontrivial
action on $G_1$ sector, we may define the action of the operator $\hat{O}_m$
to be trivial on the other sectors.
For the trivial action of the operator $\hat{O}_m$
on these sectors the corresponding matrix elements of the operator 
$\hat{O}_m$ can be set to zero. With this definition, the domain of a general
operator $\hat{O}_{4n+i}$ in the full physical solution space
is given by
\begin{equation}
\bar{D}_{4n+i} ~:=~ \bigcup_{j \neq i}G_j  ~\bigcup~ D_{4n+i} ~.
\label{lqc_dom1}
\end{equation}

We next consider the situation where the operator $\hat{O}_m$ acts non-trivially
on $G_2$, $G_3$ and $G_0$ sectors as well.
In the case of $G_1$ sector, we have seen that we can set three 
successive matrix elements of $\hat{O}_m$, equal to that of the kinematical
operator $\hat{O}$. There we have used the index of middle element 
to designate the operator itself. However the selection of three successive
matrix elements of the operator $\hat{O}_m$ in the $G_2$ and $G_3$ 
sectors can be done various ways.
So the domain of the operator $\hat{O}_m$, in these two sectors, will
depend on the particular choice one makes.
	
	A natural suggestion is to select these elements such that
the operator $\hat{O}_m$ has maximum number of successive elements that are
equal to that of the kinematical operator $\hat{O}$ around the $m$-{\it th}
element. This suggestion leads to the domain of the operator $\hat{O}_m$
in the $G_2$ sector to be $D_{m+1}$ and in the $G_3$ sector the domain could
either be $D_{m-2}$ or $D_{m+2}$.

	For the $G_0$ sector the subspace $D_{m-1}(=D_{4n})$ is trivial. So 
in this sector, we can set the matrix elements of the operator $\hat{O}_m$
as exactly same as that of the kinematical operator $\hat{O}$.
Thus we have specified all the matrix elements of the operator
$\hat{O}_m( = \hat{O}_{4n+1})$ in the full physical solution space.
Following the similar steps one can construct the operators
$\hat{O}_{4n+2}$, $\hat{O}_{4n+3}$ and $\hat{O}_{4n}$.
With this prescription, in the full physical solution space, the domain 
of a general operator $\hat{O}_{n}$ can either be chosen to be 
\begin{equation}
\bar{D}_{n} ~:=~  D_{n-1} ~\bigcup~ D_{n}
	~\bigcup~ D_{n+1} ~\bigcup~ D_{n+2} ~,
\label{lqc_dom2}
\end{equation}
or
\begin{equation}
\bar{D}_{n} ~:=~  D_{n-2} ~\bigcup~ D_{n-1}
	~\bigcup~ D_{n} ~\bigcup~ D_{n+1} ~.
\label{lqc_dom3}
\end{equation}

	For any of the above choices, the operator $\hat{O}_n$  has the same
action, as that of the kinematical operator $\hat{O}$, on at least eleven
successive basis states $|n-5\rangle, \dots |n\rangle, \dots |n+5\rangle $. 
The main difference between the two operators 
$\hat{O}_n$ and $\hat{O}$, is that the action of 
$\hat{O}_n$ on a physical state leads to another physical state whereas
the action of $\hat{O}$ does not. However they have the same action on 
the basis state $|n\rangle$ and the neighboring basis states of $|n\rangle$.

Our main aim in this exercise is to construct the Dirac observable 
corresponding to the given kinematical operator
$\hat{O}$ (\ref{lqc_oc1}). In doing this,
we have seen that it is not possible for a {\it single} Dirac observable
to represent the same action of the kinematical operator $\hat{O}$
on all the triad basis elements. Rather we need to have a set of Dirac
observables like $\hat{O}_n$. Each Dirac observable $\hat{O}_n$ mimics the 
action of the operator $\hat{O}$ only {\it on and around} the basis
state $|n\rangle$ but not globally.

	In the isotropic loop quantum cosmology, the triad basis index 
$n$ is interpreted as the {\it internal time} index. In other words, these 
discrete indices define the frame of reference for the evolution
of the universe.
Naturally the tagging of an operator by the triad index $n$ is an equivalent
description of a kind of {\it localization} of the operator.
In the section $2$, we have argued that a Dirac observable to be physically
relevant, it must be localized in an appropriate frame of reference.
Here, in the case of isotropic loop quantum cosmology, we find
a natural prescription for the {\it localization} of the Dirac observables
of the theory.

	We may define a collection of such operators $\hat{O}_n$(associated
with the same kinematical operator $\hat{O}$) as
\begin{equation}
\hat{\mathcal{O}} ~:=~ \{~ \hat{O}_{n}~;~ D(\hat{O}_{n}) 
~:=~ \bar{D}_{n} ~|~ n \in {\mathbb Z} \}~.
\label{lqc_doc}
\end{equation}
The action of any operator of the collection (\ref{lqc_doc})
on a physical state of its domain leads to a physical state.
For notational simplification we will refer this collection itself
as an operator. So symbolically its action can be given as
$|s'\rangle ~=~ \hat{\mathcal{O}}~|s\rangle$ where 
$|s\rangle$, $|s'\rangle$ both are physical states. But important point to
remember here is that $\hat{\mathcal{O}}$ is not a single operator,
rather it is a collection of operators.

	We now consider the example of volume operator in the 
isotropic loop quantum cosmology. This operator is diagonal in the triad
basis. The action of the kinematical volume operator $\hat{V}$ 
(\ref{lqc_volume_op}) on a physical state does not lead to a physical
state. So the kinematical volume operator is not a Dirac observable.
So it is necessary to construct a physical volume operator to
study its physical consequences. 

	Let us think of a situation where it is possible to construct 
a single volume operator whose action on a physical state leads to
another physical state. But without a further prescription of 
{\it localization},
this volume operator is just a {\it partial observable}. Hence this 
operator itself is not a physically relevant object. The total volume
of a dynamical universe is a evolving quantity. So it is not enough
just to say the amount of total volume, it is also necessary to specify
the corresponding time. In other words, the physical volume of the 
universe must be represented by a {\it complete observable}.

	Hence the physical volume operator should not only act
invariantly on physical solution space but it should also be {\it localized}
in an appropriate frame of reference. The localized volume operator $\hat{V}_n$,
constructed using the prescription presented here, precisely does that.
So the operator $\hat{V}_n$ is a {\it complete observable} hence it is 
a viable candidate for
the physical volume operator which represent the total volume of
the universe at a given time slice $n$. In the case of isotropic loop
quantum cosmology, the {\it localization} of the physical volume operator 
$\hat{V}_n$ is not just a choice rather a requirement of the construction.

	An important difference between the kinematical volume operator
$\hat{V}$ and the localized volume operator $\hat{V}_n$ is that the
$\hat{V}_n$ acts only on a subspace of the physical solution space 
whereas $\hat{V}$ can act on the entire physical solution space but
its action takes physical states outside the physical solution space.
However the operators $\hat{V}$ and $\hat{V}_n$ act identically on 
the basis state $|n\rangle$ and its neighboring basis states.

\subsection{Physical Inner Product}

Although there is a well defined inner product on kinematical Hilbert
space, loop quantum cosmology so far does not have an inner product defined
on physical solution space. But to obtain physically relevant 
quantities from a quantum theory, one must have some notion of 
physical inner product. So we need to define a physical inner product
for evaluation of the expectation values of the physical observables.

In this regard, we will try to learn from the reformulation of 
ordinary quantum
mechanics as a constrained dynamics. This approach is motivated by the fact
that the Hamiltonian constraint for loop quantum cosmology can be viewed
as an evolution equation \cite{mdiscrete} with respect to 
an {\it internal time}. The basic idea of {\it frozen time}
description ( see\cite{date} for a brief account and references 
therein for details) of ordinary quantum mechanics is to extend
the classical phase space by two extra dimensions of time $t$ and its
canonical conjugate $\pi_{t}$. So time $t$ is no longer a parameter of the
theory but a dynamical variable. The extended phase space is given by 
$\Gamma_{kin} := \mathbb{R}^2 \otimes \Gamma_{0}$ where 
$\Gamma_{0}$ is the usual
phase space. Consequently the kinematical Hilbert space can be given as
$\mathcal{H}_{kin} := L^2(\mathbb{R}, dt) \otimes \mathcal{H}_{0}$
where $\mathcal{H}_{0}$ is the kinematical Hilbert
space of ordinary quantum mechanics. So a general state vector
in $\mathcal{H}_{kin}$ can be written as
\begin{equation}
|\psi\rangle ~=~ \int dt ~|t\rangle \otimes |\phi_{t}\rangle 
\label{ftstate}
\end{equation}
where $|\phi_{t}\rangle$ belongs to $\mathcal{H}_{0}$. 
In this scheme one performs canonical
quantization of this new pair ($t$,$\pi_{t}$) as of an ordinary pair of
phase space variables. In ordinary quantum mechanics, physical states 
satisfy Schrodinger equation. An analogous thing is done in this 
approach by imposing the Hamiltonian constraint
on physical states $|\psi\rangle$ as
\begin{equation}
(\hat{\pi_{t}} ~+~ \hat{H}(\omega)) ~|\psi\rangle ~=~ 0 ~,
\label{ft_ham_cons}
\end{equation}
where $\hat{H}(\omega)$ is the Hamiltonian corresponding to $\mathcal{H}_0$ 
and $\omega$ denotes the phase space variables corresponding to $\Gamma_0$.

The constraint equation (\ref{ft_ham_cons}) implies that $|\phi_{t}\rangle$ has
to satisfy ordinary Schrodinger equation.  
The kinematical inner product of two physical states $|\psi\rangle$
and $|\psi'\rangle$ is given by
\begin{equation}
{\langle\psi'|\psi\rangle}_{kin} 
~=~ \int dt ~\langle\phi_{t}'|\phi_{t}\rangle ~.
\label{ft_kip}
\end{equation}
Given the fact $\langle\phi_{t}' | \phi_{t}\rangle$ is independent of time $t$,
$\langle\psi'|\psi\rangle_{kin}$ is not finite. Similar situation arises in loop
quantum cosmology when we evaluate the kinematical inner product of two physical
states. 

 	In quantum mechanics we know that the physically relevant inner
product is given in terms of inner product of $|\phi_{t}\rangle$'s.
In making the
correspondence between the inner products of two different description
of same dynamics a natural suggestions is to define a {\it physical}
inner product of $|\psi\rangle$'s as
\begin{equation}
\langle\psi'|\psi\rangle_{phys}(t) ~:=~ \langle\phi_{t}' | \phi_{t}\rangle ~. 
\label{ft_pip}
\end{equation}
One should notice that we have defined the physical inner product as 
a function of time $t$. But given the fact the evolution of $|\phi_t\rangle$'s
are unitary, the definition of physical inner product (\ref{ft_pip}) is in fact
independent of time. 
Our main aim here is to use the same idea in the loop quantum cosmology.
When recasted as a first order evolution equation \cite{mdiscrete}
in the internal time,
the corresponding evolution matrices in the loop quantum cosmology
are not unitary. 
So we will keep the definition of physical inner product tagged with time. 
In other words we have defined a family of physical inner products.

	In loop quantum cosmology a general state is expressed as 
\begin{equation}
|s\rangle ~=~ \sum_{n} s_n |n\rangle ~.
\label{lqc_state}
\end{equation}
where $\{|n\rangle\}$ is the eigen basis of the triad operator. In the 
interpretation of Hamiltonian constraint as an evolution equation
\cite{mdiscrete},
$n$ is taken to be the discrete internal time index. The kinematical
inner product of two states $|s\rangle$ and$|s'\rangle$ is given by
\begin{equation}
\langle s'|s\rangle_{kin} ~=~ \sum_{n} \langle s'|n\rangle \langle n|s\rangle~.
\label{lqc_kip}
\end{equation}

Bringing the analogy with the frozen time description of quantum mechanics
we now define a family of {\it physical} inner products of 
two states $|s\rangle$ and $|s'\rangle$, each tagged by the internal time
index $n$, as
\begin{equation}
\langle s'|s\rangle_{phys}(n) ~:=~ \langle s'|n\rangle \langle n|s\rangle ~.
\label{lqc_pip}
\end{equation}
So the physical expectation value of an operator $\hat{O}$ in the
physical state $|s\rangle$, at a given time slice $n$, can naturally be 
defined as
\begin{equation}
\langle \hat{O} \rangle_{phys}(n) ~:=~ 
{ {\langle s|\hat{O} s\rangle_{phys}(n)} 
	\over {\langle s|s \rangle_{phys}(n)} } ~.
\label{lqc_expect}
\end{equation}

\subsection{Hubble Operator}

	In the loop quantum cosmology, the {\it absence of singularity}
\cite{msingularity} is understood in two ways. Firstly the discrete
evolution equation allows the system to evolve through the internal time $n=0$ 
and secondly there exist well defined, {\it bounded} operators for the inverse 
powers of scale factor \cite{thiemannqsdV,minverse}. 
In minisuperspace quantization, however,
similar features are clearly absent and thus lead to breakdown of evolution 
equation and the curvature singularity.

	We have seen in the example of minisuperspace quantization
that the Hubble operator (\ref{ms_hubble_op}) involves the inverse power
of scale factor. So when we move into loop quantum cosmology, it is
naturally expected that the corresponding Hubble operator will have completely
different behavior near the classical singularity.

	The construction presented in the previous section for the 
{\it localized} Dirac observable is valid only for the class of operators which
are diagonal in the triad basis. The expression of the Hubble operator
(\ref{ms_hubble_op}) 
contains the connection component in its classical expression.
In the isotropic loop quantum 
cosmology, connections are represented by point holonomies which are not 
diagonal in the triad basis. So clearly we cannot apply the construction 
directly. However in the classical expression it is possible to substitute 
the term $c^2{\sqrt p}$ using the equation (\ref{ham_full1}). 

	Thereafter we can {\it define} the Hubble operator for the isotropic
loop quantum cosmology as
\begin{equation}
\hat{\mathcal{H}^2} ~=~ {k \over 3}~ \hat{p^{-{3\over 2}}}
( {\hat{H}}_{matter} ~+~{\Lambda \over k}~ \hat{p^{3\over 2}}~-~ \hat{H} ) ~.
\label{lqc_hubble_op}
\end{equation}

One should notice that the expression (\ref{lqc_hubble_op}) now has 
the total Hamiltonian term which is obviously not diagonal in the triad
basis. But the total Hamiltonian term in (\ref{lqc_hubble_op}) 
effectively drops out whenever the Hubble operator acts on the physical 
states. 

The quantization for the operator
$\hat{p^{-{3\over 2}}}$ used in \cite{minflation} has an adjustable 
ambiguity parameter. However for illustrative purpose here we will use
the simplest possible quantization for $\hat{p^{-{3\over 2}}}$. Using
the classical identity $p^{-{3\over 2}}~=~ a^{-1}~a^{-1}~a^{-1}$ we
quantize the operator as
\begin{equation}
\hat{p^{-{3\over 2}}} ~=~ {\hat{a^{-1}}}~{\hat{a^{-1}}}~{\hat{a^{-1}}} ~.
\label{lqc_p32}
\end{equation}

In the case of FRW-deSitter universe the matter sector Hamiltonian is zero.
Since the quantity $p^{3\over 2}$ is proportional to the volume, 
then using the volume operator $\hat{V}$, we can quantize the operator
$\hat{p^{3\over 2}}$ as
\begin{equation}
\hat{p^{3\over 2}} ~=~ ~{l_{p}^{-3}~\hat{V} }~.
\label{ham_matter_fds_op}
\end{equation}

We have already pointed out that the total Hamiltonian term in
(\ref{lqc_hubble_op}) effectively drops out whenever it acts on the
physical states. So for evaluating expectation values in the physical
states, we just need to consider only the rest of the terms of the
kinematical Hubble operator (\ref{lqc_hubble_op}). These terms are 
diagonal in the triad basis. 

Naturally, using the method described in the previous section for 
localized Dirac observables, we obtain the one parameter family of physical 
Hubble operators $\{\hat{\mathcal{H}^2_n}\}$, each labeled by the 
discrete time index $n$. The localized Hubble operator
$\hat{\mathcal{H}^2_n}$ is a diagonal operator and has the same action
that of kinematical Hubble operator(apart from the total Hamiltonian term)
on the basis state $|n\rangle$ and its neighboring basis states.

So for the FRW-deSitter universe the physical expectation value 
(\ref{lqc_expect}) of the Hubble operator is given by
\begin{equation}
\langle \hat{\mathcal{H}^2}\rangle_{phys}(n) 
	~=~ {\Lambda \over 3}
\left(16\gamma^{-2} l_p^{-3}~\left(\sqrt{V_{{1\over2}|n|}}
	- \sqrt{V_{{1\over2}|n|-1}}\right)^2 \right)^{3}
	~l_p^{-3} V_{{1\over2}(|n|-1)} ~.
\label{lqc_hexp_fds}
\end{equation}
For given any operator to be a viable physical observable
it must have the correct classical limit. We now show that the Hubble
operator constructed here does have the correct classical limit. 
In this context the classical limit is obtained by taking 
the large $n$ limit {\it i.e.} the large volume limit where one expects 
the effects of small scale physics are negligible.  For the large value
of $n$, the expectation value of the Hubble operator is given by
\begin{equation}
\langle \hat{\mathcal{H}^2}\rangle_{phys}(n) ~=~ {\Lambda \over 3}
	\left(1~+~{37\over16}{1\over n^2}~+~ O({1\over n^4})~\right) ~.
\label{lqc_hexp_ln}
\end{equation}
In the large volume {\it i.e.} in the large $n$ limit 
$a^2 \sim {1\over6}\gamma~n$. So the correction terms to the classical
value in (\ref{lqc_hexp_ln}) are proportional to the Barbero-Immirzi
parameter $\gamma$. In the loop quantum cosmology the discreteness
in the spectrums of geometrical operators are controlled by the
Barbero-Immirzi parameter $\gamma$.
In $\gamma \rightarrow 0$ limit {\it i.e.} when the discreteness
in the spectrums of geometrical operators disappears then the correction
terms in (\ref{lqc_hexp_ln}) also disappear and the expectation value
of the Hubble operator reduces exactly to the classical value like in 
the case of minisuperspace quantization where the geometrical operators
have continuous spectrums. This implies that the quantum correction
to the Hubble parameter in the isotropic loop quantum cosmology comes
necessarily due to the underlying discrete structure of the space.

	In fig. $2$, we have plotted the expectation value of the Hubble
operator for the FRW-deSitter universe. One should notice that near $n=0$ 
it has completely different behavior compared to the classical case.
\begin{center}
\epsfxsize=100mm
\epsfysize=70mm
\epsfbox{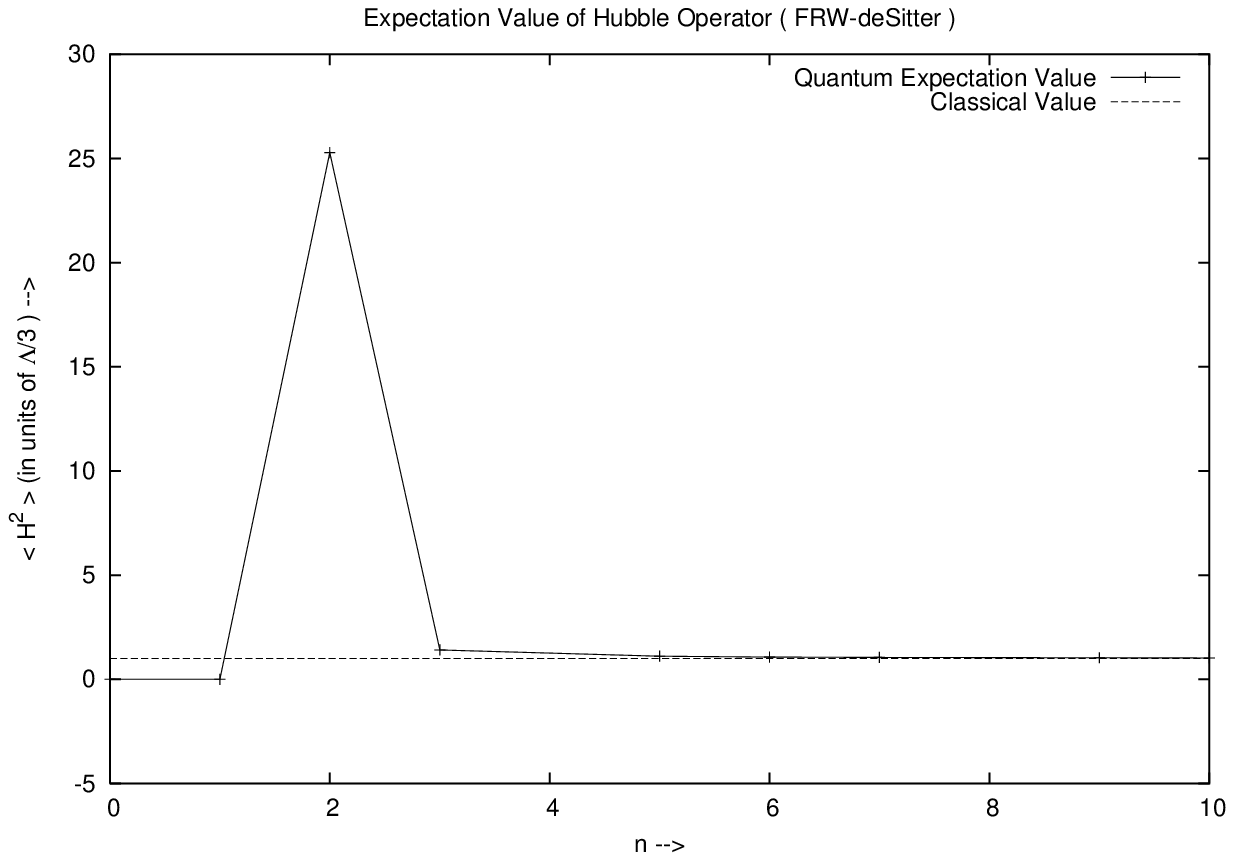}

{Figure 2. The expectation value of the Hubble operator for the 
FRW-deSitter universe}
\end{center}

We now study the case, considered in \cite{minflation}, of 
FRW universe with a conventional massless free scalar field. 
Being free field, the scalar field momentum $p_{\phi}$ is constant 
say $\omega$. So the corresponding Hamiltonian is given by
\begin{equation}
H_{matter} ~=~ {1\over 2} a^{-3}~p_{\phi}^2 
	~=~ {1\over 2}~ p^{-{3\over 2}} \omega^2 ~.
\label{ham_matter_mfs}
\end{equation}
Using the same quantization (\ref{lqc_p32}) we obtain the expectation
value of the Hubble operator as
\begin{equation}
\langle \hat{\mathcal{H}^2}\rangle_{phys}(n) 
	~=~ {k\omega^2 \over 6}
\left(16\gamma^{-2} l_p^{-3}~\left(\sqrt{V_{{1\over2}|n|}}
	- \sqrt{V_{{1\over2}|n|-1}}\right)^2 \right)^{6} ~.
\label{lqc_hexp_mfs}
\end{equation}
In fig. $3$ we have plotted the expectation value of the Hubble operator
for the FRW universe with a conventional massless free scalar field.
\begin{center}
\epsfxsize=100mm
\epsfysize=70mm
\epsfbox{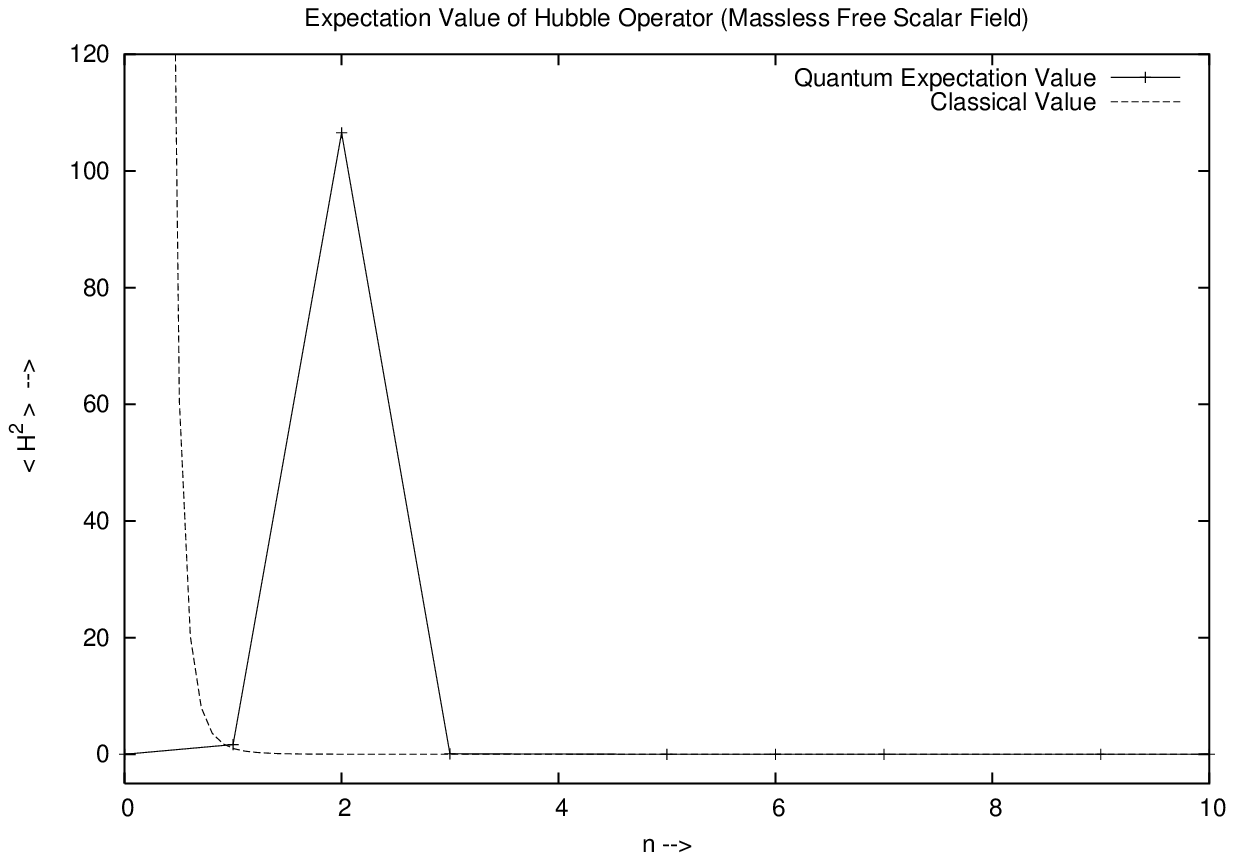}

{Figure 3. The expectation value of the Hubble operator for a FRW universe 
with a massless free scalar field}
\end{center}

So in both cases we find that there is a period of evolution where the 
expectation value of the Hubble operator increases with increasing time 
and then decreases to reach its classical value.
Classically increasing Hubble parameter square for expanding universe implies
a super-accelerating universe. In the quantum case, the expectation value
of the Hubble operator has such behavior. So one may interpret this
period as a phase of super-inflation. However, evolution being in discrete 
steps, one cannot make a direct analogy with the classical case. 
Also this period exists only for a few steps. But surely the expectation
value of the Hubble operator remains larger than the classical value for a 
period of evolution.

In \cite{minflation}, the phase of super-inflation
was obtained from the effective Friedmann equation and the reason for
this quantum geometric inflation was cited to be the modification of
the kinetic term in the matter Hamiltonian but here we find that apart
from the modification of the kinetic term, the existence of $a^{-1}$ term 
in the Hubble parameter also causes it to behave differently. 
So this phase seems to be more generic in the loop quantum cosmology.
In particular this phase exists even for the FRW-deSitter universe
where there is no matter Hamiltonian term. Naturally one gets larger
amount of inflation from this picture compared to that from the analysis of
effective Friedmann equation. It is important to emphasize here that
the consideration of different values for the ambiguity parameter in the 
quantization of $\hat{p^{-{3\over2}}}$, as used in \cite{minflation}, 
will lead to further changes in the amount of inflation.

\section{Conclusions}

	To summarize, in this paper we have emphasized the need for
{\it localization} of Dirac observables of a dynamical theory
from a general perspective.
In the case of isotropic loop quantum cosmology, we have shown that 
for a given kinematical operator, one cannot construct the corresponding
Dirac observable without {\it localization}.
Further we have presented a construction by 
which one can consider a class of operators as {\it localized} Dirac
observables. These Dirac observables can act only on a subspace 
of the space of physical solutions. Ideally one would like to have
Dirac observables that can act on the full physical solution space.
But the construction of Dirac observables here does {\it not} allow that.
However in the cosmological context it may be more reasonable to have subspaces
rather than the full solution space as the domain of Dirac observables, as
any ``observations'' in the cosmology after all would be made with a 
single state.
As an example of these Dirac observables, we have constructed the Hubble 
operator for the spatially flat isotropic universe. 

In the case of closed
isotropic model, the same construction of Dirac observable goes through
for the diagonal operators. However, using the same substitution
as for the case of flat FRW universe, the corresponding Hubble 
operator cannot be reduced to a function of triads and the total Hamiltonian
alone. In this case, the classical expression of the Hubble operator
contains an explicit connection term. So to construct the Hubble operator
for isotropic closed model, we need to have a procedure for
constructing Dirac observables for the functions of connection variables.
The corresponding operators are {\it not} diagonal in the triad basis. 
However preliminary investigation in this direction indicates such 
construction to be viable.
	
	From the physical point of view, specifying a subspace of the physical 
solution space can also be considered as fixing of some of the
initial conditions. We have seen that for a Dirac observable 
the requirement of vanishing commutator with the Hamiltonian constraint
dynamically selects out a subspace of the physical solution space.
Thus in this regard, in the isotropic loop quantum cosmology, it seems that
Dirac observables themselves supply their initial conditions.
As the selections of the subspaces depend on the particular time slices, it
also implies that the specification of the initial conditions is
{\it dynamical}.
Moreover this aspect seems to be completely different from that described
in \cite{minitial} where one selects out an unique solution by the requirement
of pre-classicality. Furthermore it seems that by combining these two aspects 
it may be possible to figure out the class of Dirac observables which are 
relevant in the large volume evolution.
	
	As yet we do not have a definition of physical inner product in the 
loop quantum cosmology, we have used here a definition of physical inner product
which is motivated from the reformulation of ordinary quantum mechanics.
But it is important to remember that this definition is a first attempt
in this context and need not be the final.
In order to define the physical inner product for the states of the 
frozen time description of the ordinary quantum mechanics, we have seen that
it is natural to define a {\it family} of physical inner products.
However
the members of this physical inner products family are not really different
from each other as the evolution in ordinary quantum mechanics is unitary.
But in the case of isotropic loop quantum cosmology the evolution
(when recasted as a first order evolution equation) in the internal 
time is not unitary. So the physical inner product family 
members are different from each other but surely they
are not completely arbitrary. In this case also the members of the physical 
inner products family are related to each others by the basic evolution 
equation as in the case of ordinary quantum mechanics.
The expectation values of the Dirac observables of the first category,
evaluated with respect to this physical inner product,
have zero uncertainties. However the physical inner product defined
here may not be general enough to be adapted in the full theory
so it is desirable that one explores other possible routes as well.

	The Hubble parameter is an extremely important quantity of the 
cosmology. So for the phenomenological purpose it is important to consider
more realistic matter Hamiltonian in evaluating the expectation values of
the Hubble operator. Apart from that one should also take care of the 
quantization ambiguities.

\section*{Acknowledgements}
I am grateful to Ghanshyam Date for careful reading of the manuscript,
further comments and suggestions on it. I am also indebted to him
for hours of useful discussions which has helped me to understand 
many ideas of this subject. I thank Martin Bojowald for providing valuable 
comments and suggestions on the preprint.
I also thank Bobby Ezhuthachan for having productive discussions.

\end{document}